# FLAMINGOS-2: The Facility Near-Infrared Wide-field Imager & Multi-Object Spectrograph for Gemini

Stephen Eikenberry, Richard Elston, S. Nicholas Raines, Jeff Julian, Kevin Hanna, David Hon, Roger Julian, R. Bandyopadhyay, J.Greg Bennett, Aaron Bessoff , Matt Branch, Richard Corley, John-David Eriksen, Skip Frommeyer, Anthony Gonzalez, Michael Herlevich, Antonio Marin-Franch, Jose Marti, Charlie Murphey, David Rashkin, Craig Warner
Department of Astronomy, University of Florida, Gainesville, FL  32611

Brian Leckie, W. Rusty Gardhouse, Murray Fletcher, Jennifer Dunn, Robert Wooff, Tim Hardy
Herzberg Institute of Astrophysics, Victoria, British Columbia

## ABSTRACT

We report on the design and status of the FLAMINGOS-2 instrument – a fully-cryogenic facility near-infrared imager and multi-object spectrograph for the Gemini 8-meter telescopes.  FLAMINGOS-2 has a refractive all-spherical optical system providing 0.18-arcsecond pixels and a 6.2-arcminute circular field-of-view on a 2048x2048-pixel HAWAII-2 0.9-2.4 μm detector array.  A slit/decker wheel mechanism allows the selection of up to 9 multi-object laser-machined plates or 3 long slits for spectroscopy over a 6x2-arcminute field of view, and selectable grisms provide resolutions from ~1300 to ~3000 over the entire spectrograph bandpass.  FLAMINGOS-2 is also compatible with the Gemini Multi-Conjugate Adaptive Optics system, providing multi-object spectroscopic capabilities over a 3x1-arcminute field with high spatial resolution (0.09-arcsec/pixel).  We review the designs of optical, mechanical, electronics, software, and On-Instrument WaveFront Sensor subsystems.  We also present the current status of the project, currently in final testing in mid-2006.

## 1. INTRODUCTION

Multi-object spectroscopy (MOS) is revolutionizing optical astronomy, in fields as far ranging as abundance studies of globular clusters to the large-scale structure of the Universe.  While MOS instruments often have somewhat lower throughput than their single-object counterparts, they overcome this loss by observing tens to hundreds of objects at a single time.  This has enabled large increases in sample sizes for many studies – often as much as 2 or more orders of magnitude.

Near-infrared spectroscopy has lagged significantly behind optical spectroscopy, with the first instruments featuring large-format (1024x1024-pixel or larger) detector arrays appearing on telescopes in just the past few years.  In particular, near-infrared MOS have only begun to appear very recently.  The first fully-cryogenic IR MOS, FLAMINGOS, was developed at the University of Florida, and has seen successful use at the Gemini South 8-m and MMT 6.5-m telescopes, and is currently in service as a facility instrument of the Kitt Peak 4-meter telescope (Elston et al., 2002).

FLAMINGOS-2 (Eikenberry et al., 2004) is a fully cryogenic near-infrared (0.9-2.4 μm) wide-field imager and multi-object spectrograph which is being built by the University of Florida Department of Astronomy for the Gemini 8-m telescopes on Mauna Kea, Hawaii and Cerro Pachon, Chile.  FLAMINGOS-2 shares much of the instrument heritage of FLAMINGOS, as both a wide-field imager and MOS.  FLAMINGOS-2 differs from FLAMINOGS primarily in having optics and opto-mechanical systems optimized for the Gemini telescopes, providing 0.18-arcsec pixels and a 6.2-arcmin field of view – covering approximately 6 times the solid angle of FLAMINGOS on the same telescopes.

In this paper, we review the design parameters for FLAMINGOS-2, and present the current status of the project. Sections 2-6 cover the design areas for the instrument (optical, mechanical, electronics, software, and the On-Instrument WaveFront Sensor subsystems).  In Section 7, we present the current status of FLAMINGOS-2, and projected dates for on-telescope commissioning.

# 2. OPTICAL SYSTEM

## 2.1 Optics Introduction

FLAMINGOS-2 is an imaging spectrometer for use at the f/16 telescope focal surface of either Gemini 8-m telescope. It consists of a collimator providing a pupil image of high quality, and a camera following the collimator to produce a reimaged focal surface on the detector array with 2048x2048 18μm pixels. A combination of filters and grisms are placed near the pupil for broad- and narrow-band imaging and moderate-resolution spectroscopy. A pupil mask reduces excess thermal emission from the telescope. The imaging mode field forms an inscribed circle on the detector. FLAMINGOS-2 may also be fed with a slower (f/30) beam provided by the Gemini Multi-Conjugate Adaptive Optics (MCAO) system. In spectroscopic mode, a selection of 9 MOS plates and 3 long slits mask off-target locations in the focal plane, passing target light through the collimator to a selectable grism inserted into the beam after the pupil. The grism disperses the incident light, which is reimaged as a spectrum on the detector array by the camera optics. We present the basic optical performance requirements for FLAMINGOS-2 in Table 1.

Table 1 – Optical Performance Requirements for FLAMINGOS-2

| Parameter | Requirement | Design Performance |
|---|---|---|
| Wavelength range | 0.9-2.5 μm | 0.9-2.5 μm |
| Input beam | Gemini f/16 | Gemini f/16 |
| Imaging field of view | ~6-arcmin circular | 6.2-arcmin circular |
| Pixel Scale | ~0.18-arcsec/pixel | 0.18-arcsec/pixel |
| Detector | HAWAII-2 (2-48x2-48 pixels, 18 μm pitch) | HAWAII-2 (2-48x2-48 pixels, 18 μm pitch) |
| Optics Throughput | >50% imaging mode<br>>30% spectroscopy mode<br>(excludes telescope, detector, atmosphere) | >75% imaging HK, >65% imaging J<br>>50% spectra HK, >40% spectra J<br>(all measured performance in lab) |
| Image Quality (50% EED) | <25 μm | <20 μm (incl. tolerances) |
| Image Quality (80% EED) | <43 μm | <40 μm (incl. tolerances) |
| Pupil Image | ~100-mm diameter | 103-mm diameter |
| Pupil Image Quality | <5-mm FWHM (2-mm goal) | <2-mm FWHM (incl. tolerances) |
| Spectral Resolution & wavelength coverage | R~1300 (2 pix; 0.9-1.8 or 1.25-2.5 ?m)<br>R~3000 (2 pix; J or H or K) | Matches requirement |
| MOS field of view | 6x2-arcmin (3x2-arcmin at full spectral range) | Matches requirement |
| Multi-conjugate Adaptive Optics mode (f/30) | 2-arcmin field of view, flat focal plane, ~50-mm pupil | Matches requirement |

## 2.2 Optical Layout

FLAMINGOS-2 uses an all-refractive (except for two fold mirrors) all-spherical optical design. The initial design work was carried out by H. Epps (UC Santa Cruz), and then modified by the FLAMINGOS-2 team and Telic Optics. The final testplated design consists of a standard collimator/camera design with 9 lenses made from two materials – $CaF_2$ and Ohara SFTM-16. We present an illustration of this design in Figure 1. Light enters from the telescope at the left, coming to a focus approximately 300-mm behind the 280-mm-diameter $CaF_2$ entrance window. The all-$CaF_2$ collimator consists of the large (270-mm diameter) singlet, followed by a closely-spaced pair of lenses, separated from the first lens by a pair of space-saving fold mirrors. Light then passes through a bandpass filter before forming a 102-mm diameter pupil image at the cold stop. It then passes through a 6-lens camera comprised of what are essentially two airspaced triplets. Both camera triplets contain two lenses of $CaF_2$ and one of SFTM-16 (middle lens of triplet 1 and last lens of triplet 2). Light finally comes to a focus on the 2048x2048-pixel HAWAII-2 detector array at the right. Grisms can be inserted in the beam in the collimated airspace between the pupil stop and the first camera lens. The total track length of the optical axis, from entrance window to detector array, is 2.5-meters.

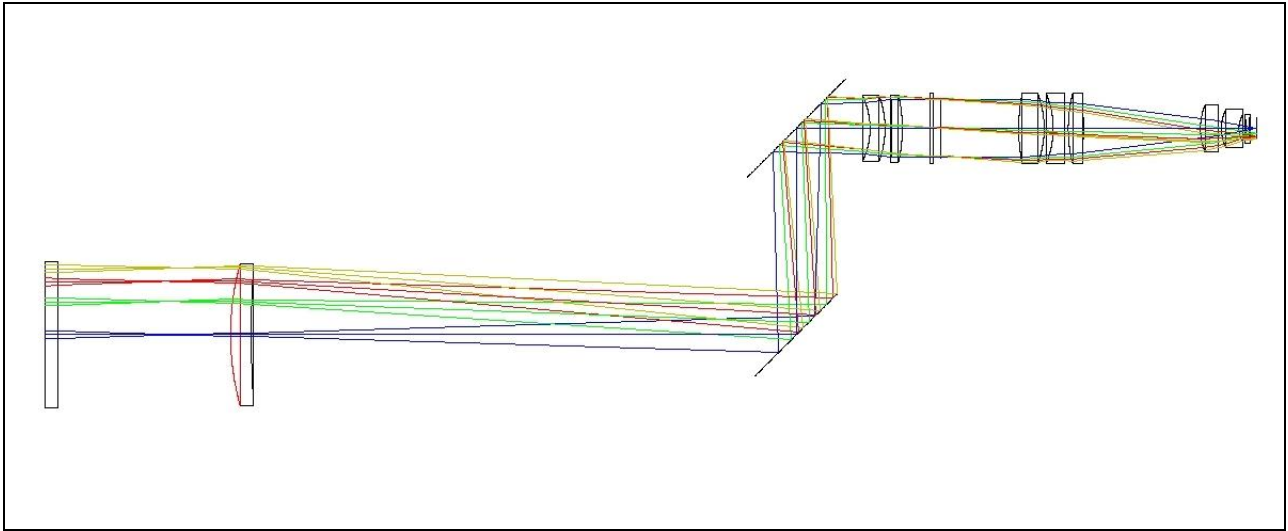

**Figure 1** – FLAMINGOS-2 optical layout.

SFTM-16 is a common optical glass, but seldom used in the infrared. A consequence of this situation is that we could find no information on the cryogenic indices of refraction for SFTM-16. For FLAMINGOS-2, we obtained a large sample of SFTM-16, from which we cut and polished prisms for use in cryogenic index measurements, in collaboration with W. Brown (CfA) and the Optical Sciences Center at the University of Arizona (Brown et al., 2004). Through the use of CaF2 with SFTM-16, we obtain very good performance in image quality and achromaticity over a wide field of view, without needing to resort to higher-index or more difficult materials such as ZnSe, ZnS, or $BaF_2$.

### 2.3 Expected Performance

The modeled performance of the FLAMINGOS-2 optical system, including standard fabrication and alignment tolerances, meets or exceeds all specifications shown in Table 1. As an example, in Figures 2-3, we show K-band spot diagrams and encircled energy curves for a simulated FLAMINGOS-2 optical system which embodies the 90%-ile "worst case" scenario, based on a Monte Carlo simulation of the specified tolerances for all optical components. We note that at this writing, all optics have been delivered, integrated, and aligned within all specifications.

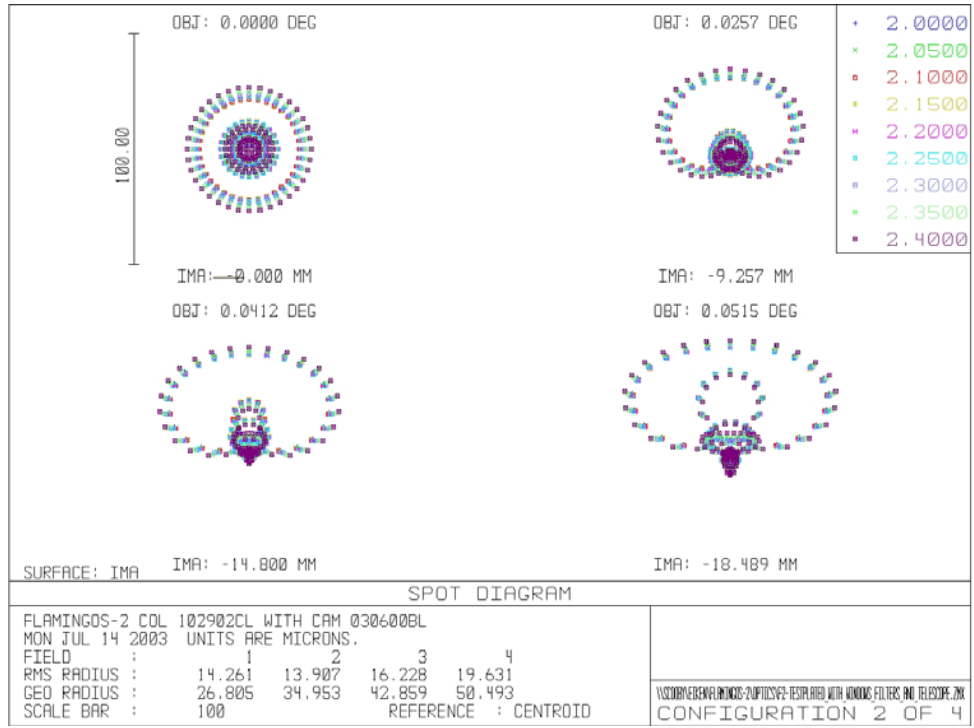

**Figure 2** – K-band spot diagram for FLAMINGOS-2 (90th percentile "worst case" realization based on Monte Carlo simulations of nominal fabrication and alignment tolerances).

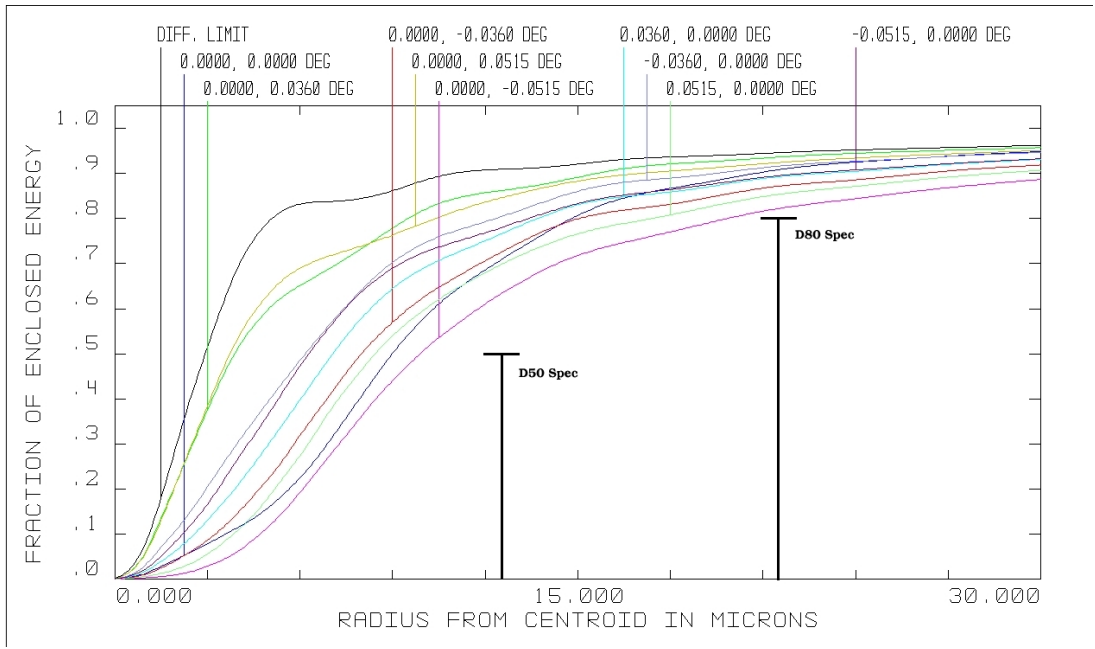

**Figure 3** – K-band encircled energy diagram for FLAMINGOS-2 (90th percentile "worst case" realization based on Monte Carlo simulations of nominal fabrication and alignment tolerances). Vertical bars indicate requirements for 50% and 80% encircled energy.

# 3. MECHANICAL SYSTEM

## 3.1 Mechanical Overview

FLAMINGOS-2 consists of two vacuum dewars connected by mechanical and vacuum adapters, both cooled by cryocooling systems (Figure 4). The smaller "MOS" dewar contains the MOS/Decker wheel mechanisms, including the slit plates and imaging field stops. It is connected to the larger "Main" dewar via a vacuum bellows and tapering vacuum adapter, with an ambient-temperature gate valve allowing the two dewars to be separately vacuum- and thermally-cycled.

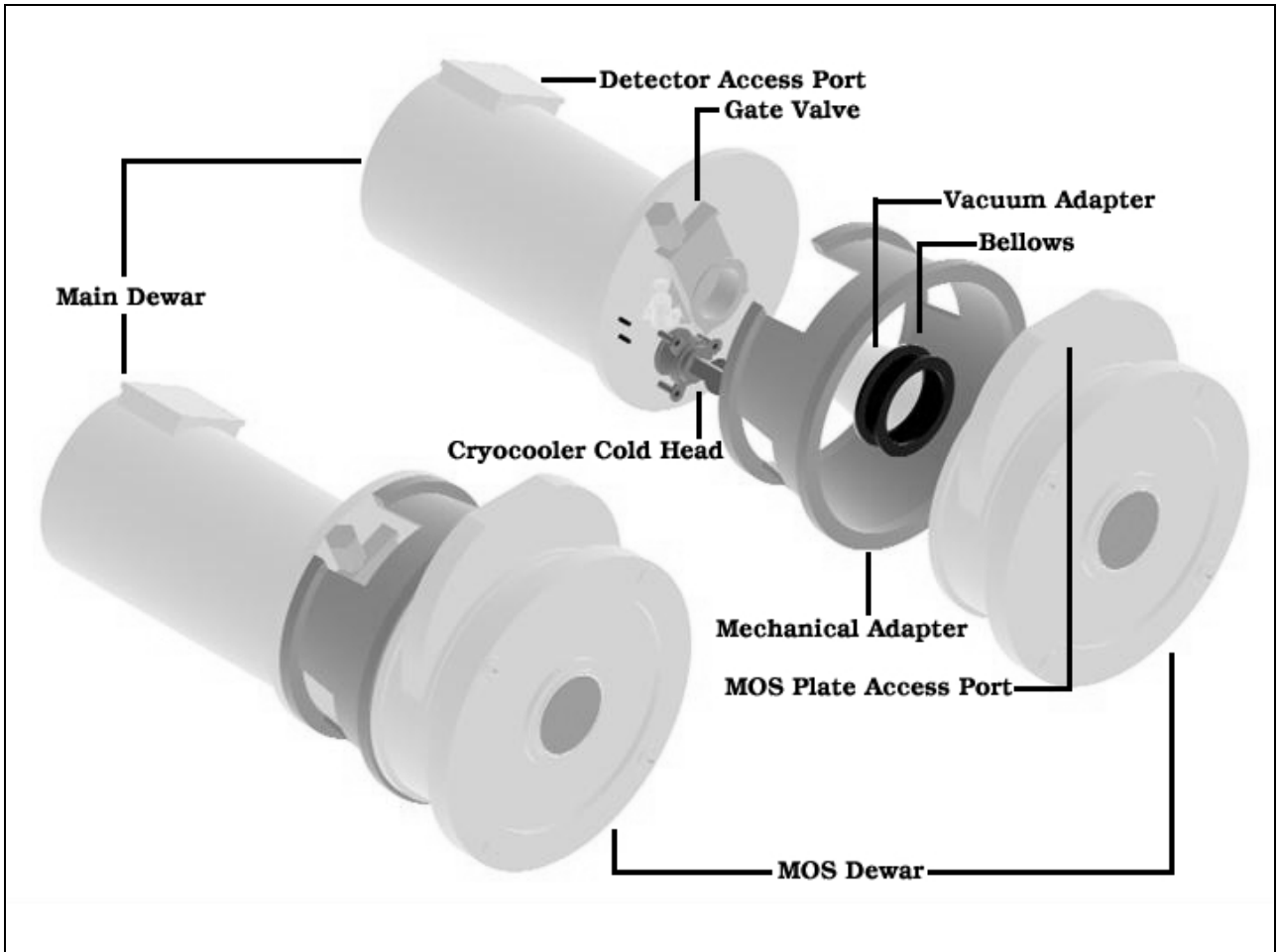

**Figure 4** – Overview mechanical layout of FLAMINGOS-2

## 3.2 MOS Dewar

The FLAMINGOS-2 MOS dewar includes the interface to the Gemini Instrument Support System (ISS), and contains the dewar entrance window, the On-Instrument WaveFront Sensor (OIWFS) subsystem (see below), MOS/Decker wheel mechanisms, and first collimator lens. The MOS wheel is driven by a worm gear and has detents for each position, including 9 MOS plates, 3 long slits, and 3 imaging field stops. It is immediately followed by a 3-position Decker wheel (one position each for MOS, long-slit, and imaging stops). All wheel mechanisms in FLAMINGOS-2 are driven by the same cryogenically-modified Portescap motors used by our team in FLAMINGOS and the Gemini facility mid-infrared instrument T-ReCS, and use identical cryogenic microswitches.

The MOS dewar is cooled by a CTI Cryogenics model 1050 cryocooler, to an operating temperature of 100K. The MOS dewar is designed to be thermally- and vacuum-cycled separately from the main dewar, allowing daytime changeouts of the MOS plates via the MOS plate access port. The required warm up time for the MOS dewar is 4 hours, aided by a bank of warm-up heaters. The required cooldown time to operating temperature is 6 hours. Lakeshore diodes are used as temperature sensors throughout the MOS dewar.

### 3.3 Main Dewar

The FLAMINGOS-2 main dewar contains the remainder of the collimator optics (two fold mirrors and 2 lenses), the filter, Lyot, and grism mechanisms, the camera optics, and the HAWAII-2 detector array. The internal layout of the main dewar is shown in Figure 5. The large circular "lid" of the dewar is at room temperature, and is separated from the main optical bench by a G10 fiberglass "can". All remaining opto-mechanical components of FLAMINGOS-2 attach to the optical bench, which is a lightweighted aluminum structure with honeycombed pockets milled on its underside. We added a semi-cylindrical "pig trough" for additional mechanical rigidity after finite element analyses by our team and Quartus Engineering (San Diego) revealed that the originally-designed gusseted structure would fail to meet flexure specifications. Figure 6 provides an exploded view of the filter/Lyot/grism wheel box, including 2 filter wheels, a Lyot wheel (including selectable pupil stops for the seeing-limited f/16 mode and the MCAO f/30 mode), and a grism wheel. All wheels have 5 positions, cryogenic microswitch homing sensors, and are driven by cryogenically-modified Portescap motors.

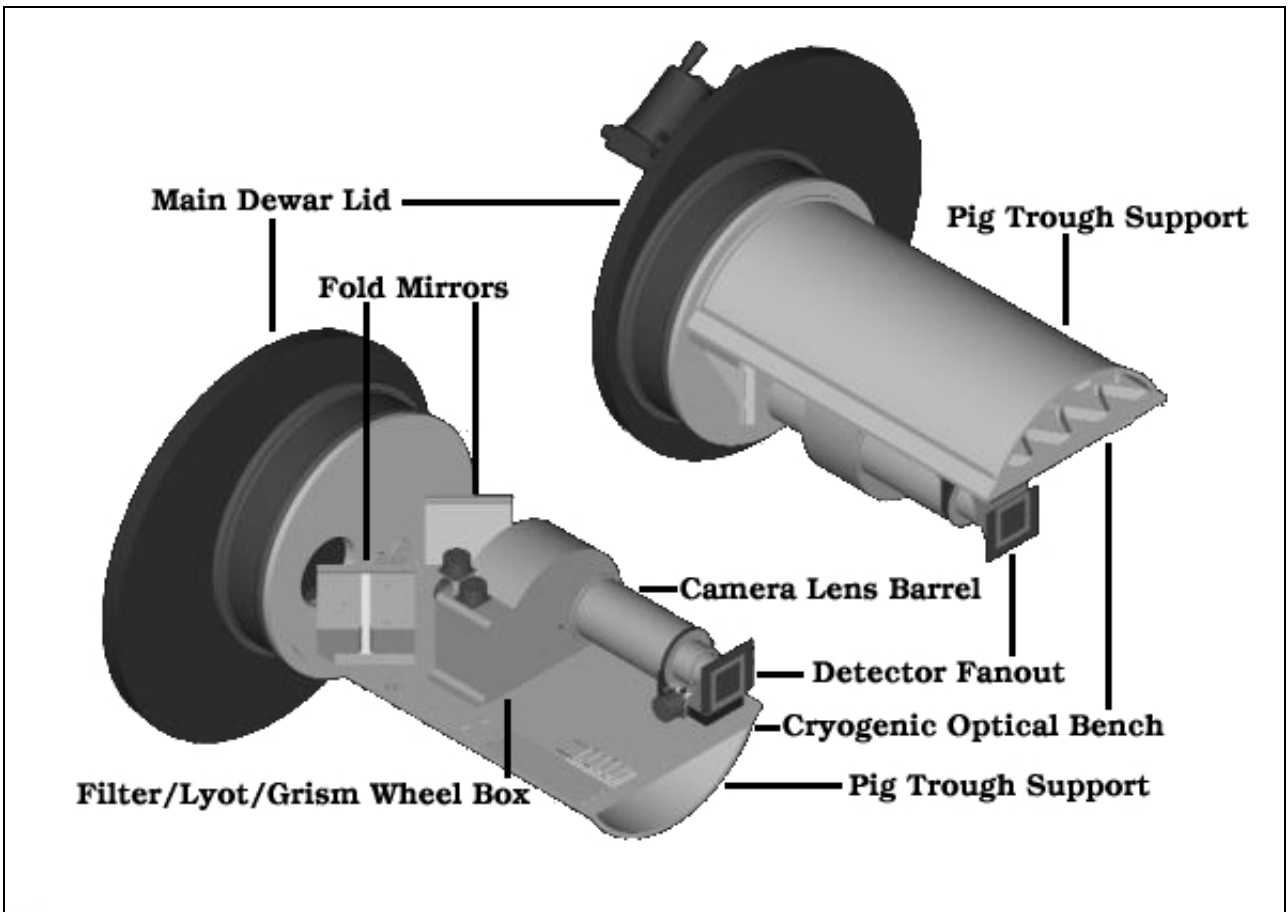

**Figure 5** – Internal layout of the main FLAMINGOS-2 dewar.

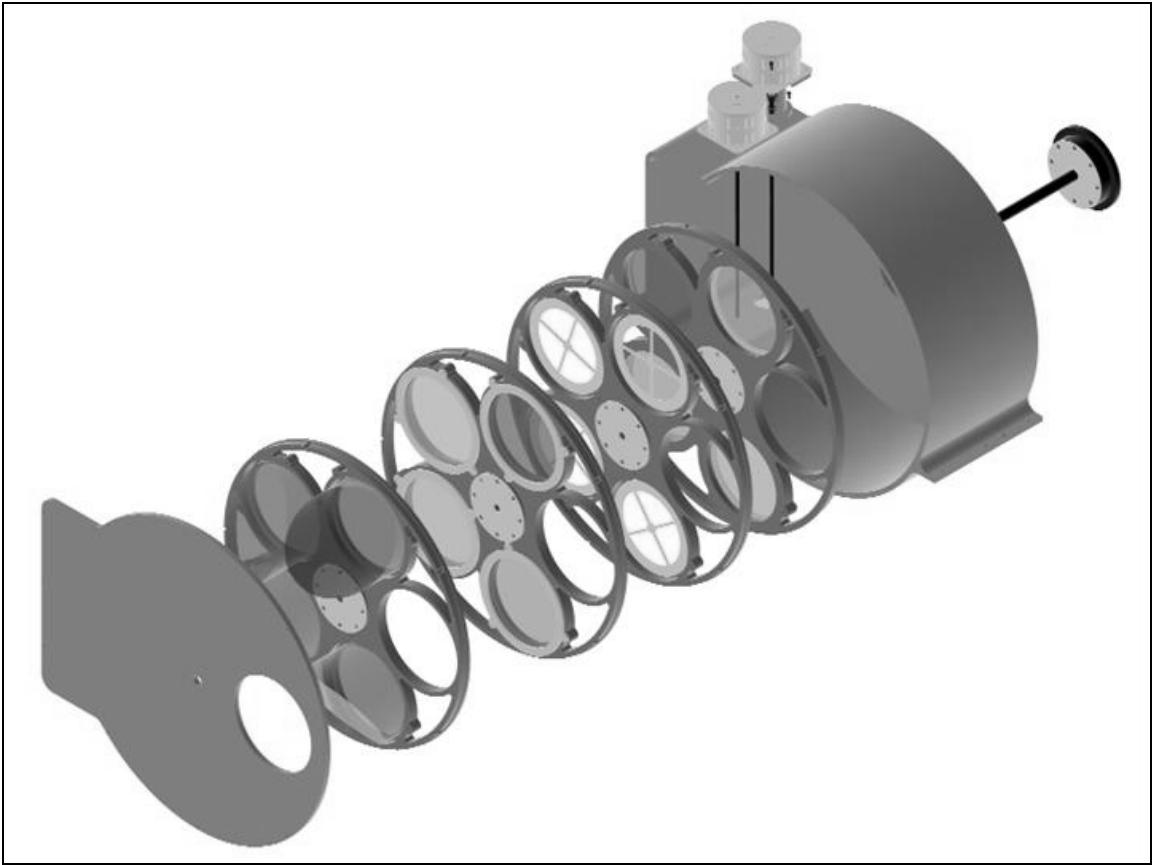

**Figure 6** – Exploded view of the filter/Lyot/grism wheel box, including 2 filter wheels, a Lyot wheel, and grism wheel.

Following the wheel box is a single camera barrel, which holds the 6 camera lenses. Immediately after the camera barrel is the detector mount, which includes an LVDT-monitored adjustable focus positioner.

The main dewar is cooled by a CTI Cryogenics model 1050 cryocooler, identical to the one used in the MOS dewar. Cold strapping to the cold head keeps the detector at a selectable temperature in the 60-80K range, while the remainder of the main dewar components are held at 100K. The temperature of various portions of the main dewar are monitored by Lakeshore diodes, and the detector array has a separate thermal control loop, including a Lakeshore diode sensor and a low-power heater resistor for temperature stabilization. Two banks of higher-powered heater resistors attached to the optical bench provide warm-up assistance.

## 4. ELECTRONICS

The electronics systems of FLAMINGOS-2 are closely based on the electronics successfully used in the FLAMINGOS instrument. We use all 32 outputs of the HAWAII-2 array for readout. FLAMINGOS-2 uses the same MCE4 array controller developed at the University of Florida and GATIR by Kevin Hanna for FLAMINGOS and T-ReCS. The basic features of this controller are described in Elston et al. (2002), and we refer the reader to that paper for details. The primary difference between the MCE4 array controller for FLAMINGOS-2 and the description of Elston et al. (2002) is that the original FLAMINGOS MCE4 controller has been upgraded to use fully-differential lines in its pre-amplification stage. This removes a large amount of pickup noise seen in the FLAMINGOS detector test data, and reduces the effective read noise by ~35% for a single read. With multiple sampling, the effective read noise has been demonstrated at <5 e- RMS, and will provide background-limited performance for FLAMINGOS-2 even for short imaging exposures in the low-background J-band.

## 5. SOFTWARE

The FLAMINGOS-2 instrument software system is required to command and control the instrument (including mechanisms, thermal systems, etc.) and handle data acquisition, as well as interface to an engineering control console, and the observatory software system. The instrument control is handled via the UFLIB protocols used for FLAMINGOS and T-ReCS, with individual "agents" handling each controllable device in the instrument. These agents can be controlled via PERL scripts for low-level laboratory testing and operation. For actual observatory operations, the agents are controlled via a central EPICS database. The EPICS database has two primary levels – a lower-level device control layer, and a higher-level Instrument Sequencer (IS) layer. The IS handles all communication with the observatory software system. This architecture is very similar to that used for T-ReCS, and in fact ~75% of the code for FLAMINGOS-2 is directly copied from T-ReCS. The remaining modifications primarily reflect the different number and types of devices in the two instruments, and the rather fundamental differences in observing modes between mid- and near-infrared instruments.

In addition to the control system, the University of Florida instrument team has developed a data reduction pipeline for analyzing both laboratory and on-telescope data from FLAMINGOS-2. The Florida Analysis Tool Born of Yearning for high-quality scientific data (FATBOY) runs as a Python/Pyraf-based application for imaging, spectroscopy, and multi-object spectroscopy data reduction. FATBOY is currently undergoing beta-testing using both F2 laboratory results as well as on-telescope data from FLAMINGOS.

## 6. ON-INSTRUMENT WAVEFRONT SENSOR

The On-Instrument WaveFront Sensor (OIWFS) for FLAMINGOS-2 provides fast tip-tilt guiding and focus information to the acquisition and guiding system of the Gemini telescope. The FLAMINGOS-2 OIWFS subsystem was built by the Herzberg Institute of Astrophysics (HIA), in collaboration with the University of Florida, and is closely-based on the HIA-built OIWFS subsystems for GMOS. As the Gemini telescope tracks across the sky, the field of view at the input focus of the FLAMINGOS-2 instrument may drift slightly and change focus due to flexures associated with the telescope optical and mechanical system as well as the instrument support itself. There may also be slight tracking errors in the pointing of the telescope, telescope windshake and atmospheric image motion. All of these can result in a degradation of the images or throughput at a slit. To minimize these effects the OIWFS monitors the position and focus of a guide star in the field of view. Any errors detected can be 'reported' to the telescope control system and immediately corrected. Experience shows that the delivered image quality and throughput through apertures such as FLAMINGOS-2 MOS slits are significantly improved, especially for long exposures. Since the OIWFS is located in the MOS dewar, it also provides an efficient and effective way to precisely acquire fields.

The FLAMINGOS-2 OIWFS probe patrols the bulk of the FLAMINGOS-2 FOV. The portion of the probe arm that enters the FLAMINGOS-2 field-of-view is cooled, to avoid contaminating the infrared images. The FLAMINGOS-2 OIWFS uses the same CCD and CCD controller as was used on GMOS.

When the probe tip has been centered on a guide star the light passes through a weak positive lens which slightly modifies the f/ratio of the beam to be fed to the remaining optics. Following the lens is a mirror which folds the converging beam through approximately 90 degrees towards a field stop. An image of the guide star is formed at the field stop. From the field stop the now diverging beam passes to a collimator lens. The collimated beam proceeds through a filter to the Shack-Hartmann 2x2 lenslet array which is located at an image of the telescope exit pupil. The lenslet array 'dices' the pupil patch into four converging beams which are intercepted by a re-imaging lens that forms four images of the guide star on a CCD. The positions of the four images can be used to monitor the guide star position and any focus changes that occur. Image motion is manifested as synchronized motions of the four guide star images. Changes in focus result in the images moving along diagonal paths in synchronized radial motions.

Mechanically, the FLAMINGOS-2 OIWFS has been modified from GMOS to be compatible to a moderate vacuum environment with the probe arm operating at <140K. In addition, the OIWFS has been modified to fit between the Flamingos 2 MOS lid and the outer radiation shield of the mask wheel. We show the resulting design in Figure 7. As with the GMOS OIWFS, the FLAMINGOS-2 OIWFS is composed of two stacked rotational stages, each supported by a single, preloaded, X-contact bearing. The top stage supports the optics package which includes the lenslet array, pickoff arm and CCD. The goal of the OIWFS design is to minimize differential flexure over the observing time period, rather than absolute flexure of the structure. Both stages are driven by stepper motors through a worm gear system, with ratios of 360:1 for the base stage and 180:1 for the pickoff stage and optics. Each stage is preloaded with a constant tension to reduce gear backlash. The position of each stage is held by a power off brake on each motor, minimizing the heat dissipation of the OIWFS. The motors, identical to the ones on GMOS, achieve measured RMS position repeatability of <35-mas projected on the sky. In this design, the CCD's temperature is reduced by cold strapping instead of glycol cooling, allowing a lightweight, self-supporting Kapton cable to connect the CCD to it to its bulkhead fitting. All other cabling is routed through the base stage's center of rotation to the appropriate bulkhead connections.

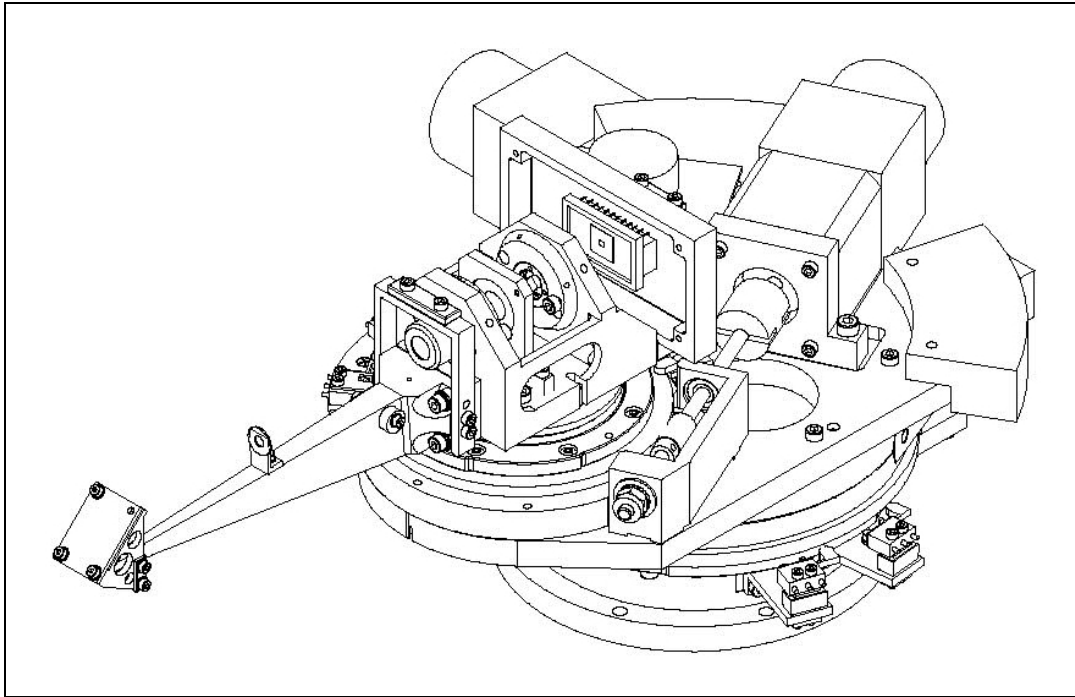

**Figure 7** – FLAMINGOS-2 OIWFS layout. The telescope optical axis is "vertical", with the telescope beam entering from the bottom of the illustration, and the MOS wheel just "above" the OIWFS probe arm.

Other important modifications to the GMOS OIWFS design were required. The GMOS OIWFS had each stage balanced in torque and moment in a cascading manner. This resulted in a long shaft extending from the pickoff stage to support the counterweight. In the current space constraints, this long counterweight shaft was not possible. With the removal of the shaft, every effort was taken to reduce both the weight and overall height of the OIWFS, accomplished by simplifying the design wherever possible. The rationalization of the design minimizes the moment acting on each bearing. Thus the weight of the OIWFS has been reduced from 14.0 kg to 5.3 kg, while retaining the torque balance of the pickoff stage. Spring motors countering the residual torque of each stage have been retained. The bearings are now loaded above their plane. In the case of the bearing supporting the pickoff arm and optics, the 0.90 kg load is 37.0 mm above the plane of the bearing. The bearing supporting the pickoff stage carries its 3.9 kg load 35.3 mm above the bearing's plane. The unobstructed range of motion of the FLAMINGOS-2 OIWFS is greater than that of the GMOS OIWFS. This allowed the replacement of the two resolvers and their mounting hardware with four micro switches per stage. Not only did this reduce both weight and complexity, it allowed a significant reduction in the height of the OIWFS. As noted earlier, the pickoff arm and its optics are required to operate at <140K. Due to the thermal expansion considerations the FLAMINGOS-2 OIWFS uses a fold mirror to capture the star instead of a prism. As an added benefit, since a mirror can steer a beam in any direction, only two angles are needed to ensure the instrument always points at the center of the secondary mirror. These angles are carried on the pickoff stage mount and the base mount. If required, these simple parts can be easily removed for re-machining the angle. The material chosen for the pickoff arm is annealed A2 tool steel. Even though the arm will not be hardened, this free machining material has better dimensional stability both during and after machining than mild steel (AISI 1015).

As of Spring 2006, the FLAMINGOS-2 OIWFS has been delivered to the University of Florida and integrated and tested in the MOS dewar.

## 7. FLAMINGOS-2 PROJECT STATUS

As of January 2006, FLAMINGOS-2 was fully assembled and integrated, and is currently undergoing laboratory checkout and testing. FLAMINGOS-2 is currently scheduled to complete pre-ship acceptance testing in early fall 2006. At that point, the instrument will be shipped to Gemini South for final acceptance testing and commissioning in its f/16 seeing-limited mode. At a later date, after the commissioning of the Gemini MCAO system, FLAMINGOS-2 will be commissioned as an MCAO MOS instrument as well.

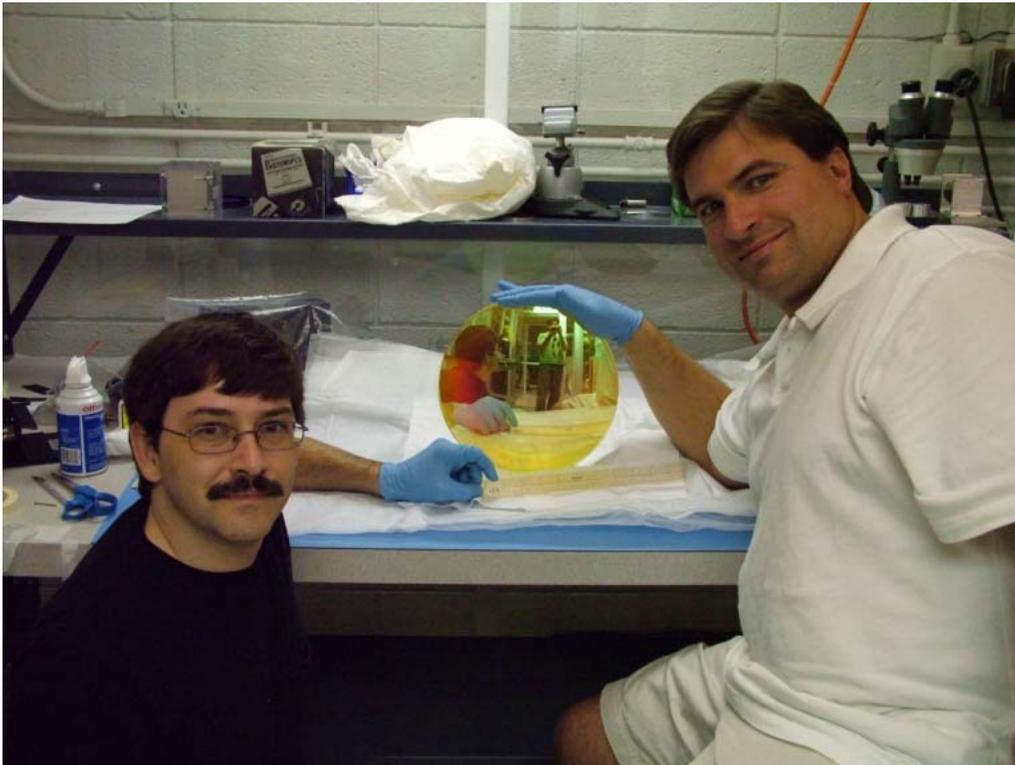

**Figure 8** – FLAMINGOS-2 Instrument Scientist S.N. Raines (left) and Principal Investigator S. Eikenberry (right) with the 270-mm-diameter $CaF_2$ collimator field lens.

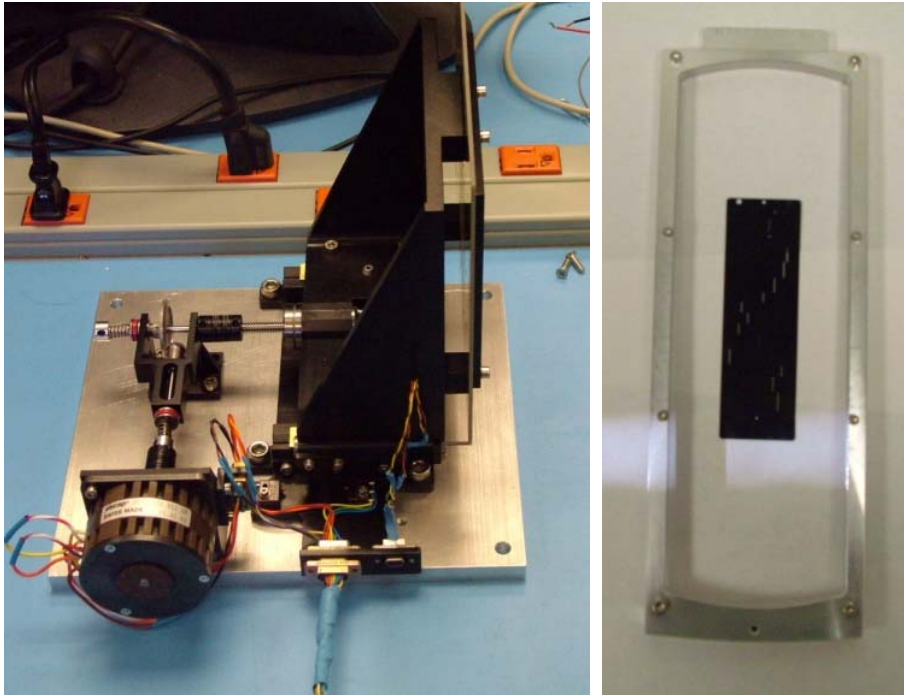

Figure 9 – (Left) Detector focus stage assembly; (Right) FLAMINGOS-2 MOS mask frame (before anodizing) with inset FLAMINGOS MOS mask for size comparison.

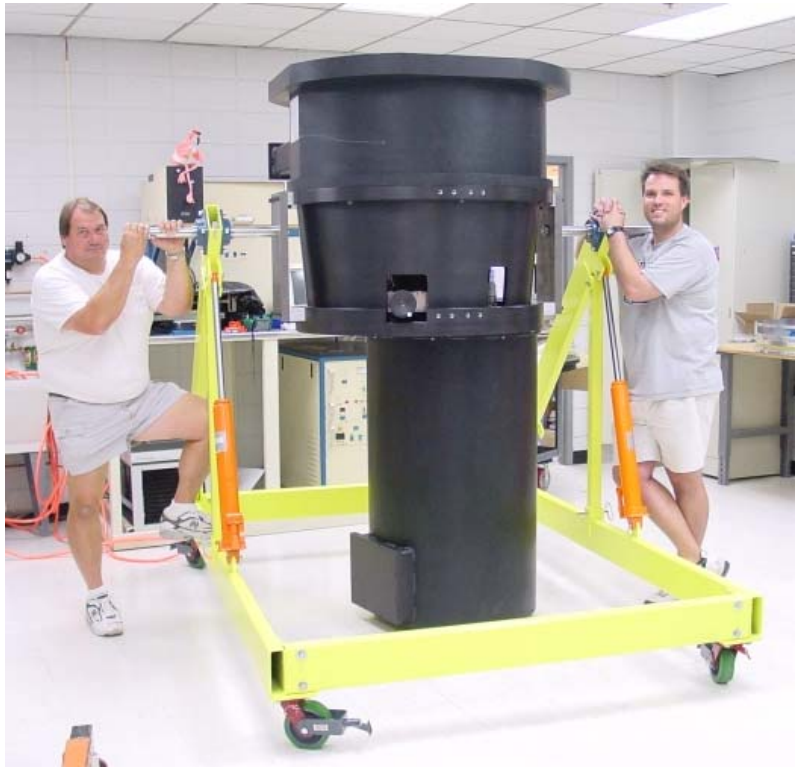

Figure 10 – FLAMINGOS-2 assembled and on its handling cart with mechanical engineers G. Bennett (left) and J. Julian (right).

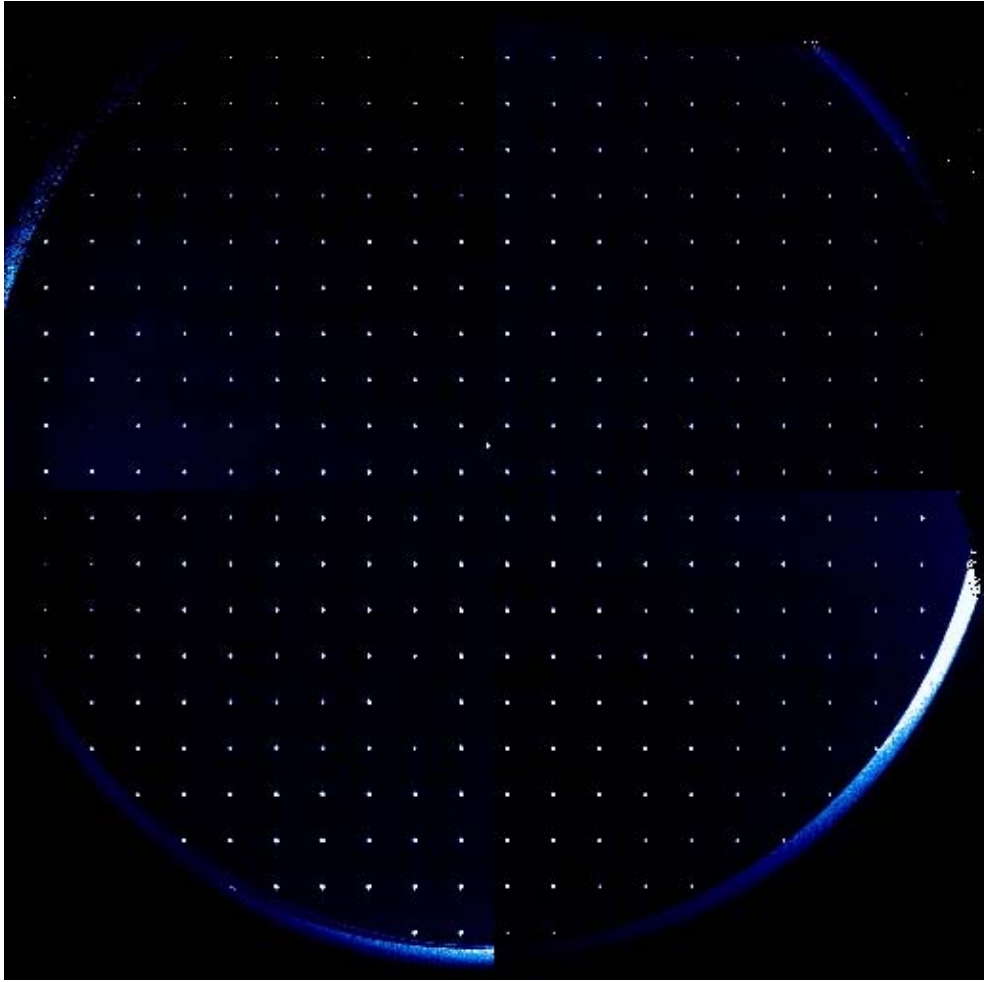

Figure 11 – Laboratory "first light" image with FLAMINGOS-2. This is a short exposure of a pinhole mask in the MOS wheel imaging stop location in the H-band.